\documentclass[prd,superscriptaddress,showpacs,showkeys]{revtex4-1}

\usepackage{listings}
\usepackage{mathtools}
\usepackage{amsfonts}
\usepackage{amsthm}
\usepackage{amssymb}
\usepackage{eurosym}
\usepackage{amsmath}
\usepackage{epsfig}
\usepackage{graphics}
\usepackage{color}
\usepackage{graphicx}
\usepackage[font={footnotesize,it}]{caption}
\usepackage[colorlinks=true,
            linkcolor=blue,
            urlcolor=blue,
            citecolor=blue]{hyperref}

\makeindex

\begin{document}

\title{Particle collisions near a three-dimensional warped AdS black hole}
\author{Ram\'{o}n B\'{e}car}
\email{rbecar@uct.cl}
\affiliation{Departamento de Ciencias Matem\'{a}ticas y F\'{\i}sicas, Universidad Cat\'{o}%
lica de Temuco, Montt 56, Casilla 15-D, Temuco, Chile}
\author{P. A. Gonz\'{a}lez}
\email{pablo.gonzalez@udp.cl}
\affiliation{Facultad de Ingenier\'{\i}a y Ciencias, Universidad Diego Portales, Avenida Ej\'{e}%
rcito Libertador 441, Casilla 298-V, Santiago, Chile.}
\author{Yerko V\'{a}squez}
\email{yvasquez@userena.cl}
\affiliation{Departamento de F\'{\i}sica y Astronom\'{\i}a, Facultad de Ciencias, Universidad de La Serena,\\
Avenida Cisternas 1200, La Serena, Chile.}
\date{\today}

\begin{abstract}

In this paper we consider the warped AdS$_{3}$ black hole solution of 
topologically massive gravity 
with a negative cosmological constant, 
and we investigate the possibility that it 
acts as a particle accelerator by analyzing  
the energy in the center of mass (CM) frame of two colliding particles in the vicinity of its horizon, 
which is known as the Ba\~nados, Silk and West (BSW) process. Mainly, we show that the critical angular momentum $(L_c)$ of the particle decreases when the parameter that controls the stretching deformation ($\nu$) increases. 
Also, we show that despite the particle with $L_c$ can exist for certain values of the conserved energy outside 
the horizon, it will never reach the event horizon; therefore, the black hole can not act as a particle accelerator with arbitrarily high CM energy on the event horizon. However, such 
particle 
could also exist inside the outer horizon 
being the BSW process possible on the inner horizon. On the other hand,  
for the extremal warped AdS$_{3}$ black hole, the particle with $L_c$ and energy $E$ could exist outside the event horizon and the CM energy blows up 
on the event horizon
if 
its conserved energy fulfill the condition $E^{2}>\frac{(\nu^{2}+3)l^{2}}{3(\nu^{2}-1)}$, 
being the BSW process possible. 

\end{abstract}

\maketitle


\tableofcontents


\newpage
\section{Introduction}

Ba\~nados, Silk and West (BSW) \cite{Banados:2009pr}  demonstrated some years ago that two particles colliding near the degenerate horizon of an extreme Kerr black hole could create large center of mass (CM) energy if one of the particles has a critical angular momentum; thus, extreme Kerr black holes can act as natural particle accelerators. Nowadays, this process is known as the BSW mechanism. 
Then, based on that the infinite acceleration can occur not only for extremal black holes 
but also for non-extremal ones \cite{Grib:2010at}, it was shown that 
an infinite energy in the CM frame of colliding particles is an universal property of rotating black holes as long as the angular momentum of one of the colliding particles approaches to the critical value \cite{Zaslavskii:2010jd} . The requirement that the black hole be rotating seems to be a essential ingredient to obtain ultra-high CM energy; however, 
it was shown that the similar effect exist for non-rotating charged black holes 
\cite{Zaslavskii:2010aw}. Also, the extension of the BSW mechanism to non-extremal backgrounds shows that particles can not collide with arbitrarily high energies at the outer horizon and that ultra energetic collisions only can occur near the Cauchy horizon of a Kerr black hole with any spin parameter \cite{Gao:2011sv}.
The non-extremal Kerr-de Sitter black holes could also act as particle accelerators with arbitrarily high CM energy if one of the colliding particles has the critical angular momentum \cite{Li:2010ej}.\\

The BSW mechanism has been studied for different black hole geometries, for instance, 
rotating charged cylindrical black holes 
was studied 
in Ref. \cite{Said:2011qm} and an extreme rotating black holes in Horava-Lifshitz gravity 
in Ref.
\cite{Abdujabbarov:2011af};
however, 
the fundamental parameter of Horava-Lifshitz gravity avoids obtaining an infinite value of the CM energy. Higher-dimensional 
rotating black holes 
have been studied 
in Refs. \cite{Abdujabbarov:2013qka, Debnath:2015bna, Zaslavskii:2016stw} and lower-dimensional black holes 
in Refs. \cite{Sadeghi:2013gmf, Lake:2010bq, Yang:2012we, Tsukamoto:2017rrl,Fernando:2017kut}. 
The particle collisions near the cosmological horizon of non-extremal Reissner-Nordstrom de Sitter black holes was studied in Ref. \cite{Zhong:2011vq}, charged dilatonic black holes in Ref. \cite{Pradhan:2014eza}, 
no-rotating  
and rotating regular black holes in Refs. \cite{Pradhan:2014oaa,Ghosh:2014mea, Ghosh:2015pra, Amir:2015pja}, 
and extremal 
modified Hayward and Bardeen rotating 
black holes in Ref. \cite{Pourhassan:2015lfa}.
On the other hand, 
charged particles in general stationary charged black holes was considered in Ref. \cite{Yi:2011}, and on  
string black holes 
in Ref. \cite{Fernando:2013qba}. The collisions of spinning particles on rotating black holes 
in Ref. \cite{Guo:2016vbt,Zhang:2016btg}, and on 
Schwarzschild black holes was considered in Ref. \cite{Armaza:2015eha}; however, the
unavoidable appearance of superluminal motion and the change of trajectories from timelike to spacelike 
can be avoided due to
the energy in the CM frame can grow unbounded provided that one of the particles is not exactly critical but slightly deviates from the critical trajectory \cite{Zaslavskii:2016dfh}. \\

The aim of this work is to consider the special class of well-known three-dimensional warped anti-de Siter black holes solutions and to study, via the BSW mechanism, the possibility to obtain unbounded energy in the CM frame of two colliding particles 
and to analyze the effect of the warped parameter that controls the stretching deformation on this. 
It is worth noting that the warped AdS$_{3}$ black holes can be viewed as discrete quotients of warped AdS$_{3}$ spacetime just like the BTZ black holes as discrete quotients of the AdS$_{3}$. Also, when the warped parameter $\nu=1$ the metric reduces to the metric of BTZ black hole in a rotating frame. An important feature of these black holes is that the Killing vector $\partial_{t}$ is spacelike everywhere in the spacetime and consequently its ergoregion extends to infinity. Also,
it is known that 
two particles in the ergosphere lead to infinity growth of the energy of the CM frame, provided the angular momentum of one of the two particles have large negative angular momentum and a fixed energy at infinity for the Kerr black holes \cite{Grib:2010at}  
which subsequently was proved to be a  
universal property of the ergosphere \cite{Zaslavsky:2013dra}.\\

It worth to mention that the collision of two particles near the horizon of a BTZ black hole was studied in Ref. \cite{Lake:2010bq, Yang:2012we, Tsukamoto:2017rrl}. In Ref. \cite{Lake:2010bq,Yang:2012we} the authors found that the particle with the critical angular momentum could exist inside the outer horizon of the BTZ black hole regardless of the particle energy being the BSW process possible on the inner horizon for the non-extremal BTZ black hole. Also, the BSW process could also happen for the extremal BTZ black hole, where the particle with the critical angular momentum could only exist on the degenerate horizon. On the other hand, in Ref. \cite{Tsukamoto:2017rrl}, the authors have studied  the collision of two particles on an event horizon and outside of the BTZ black hole, and they have shown that if well in principle the CM energy of two ingoing particles can be arbitrarily large on an event horizon if either of the two particles has a critical angular momentum and the other has a non-critical angular momentum, the critical particles never reach the event horizon. However, the motion of a particle with a subcritical angular momentum is allowed near an extremal rotating BTZ black hole and that a CM energy for a tail-on collision at a point can be arbitrarily large in a critical angular momentum limit. \\

The manuscript is organized as follows: In Sec. \ref{background} we give a brief review of the three-dimensional warped AdS black hole. Then, we study the particles motion in the three-dimensional warped black hole background in Sec. \ref{motion}. In Sec. \ref{CMS} we obtain the CM energy of two colliding particles, and in Sec. \ref{two} we study the radial motion of a particle with critical angular momentum and we investigate the possibility that
the black hole acts as a particle accelerator.  Finally, our conclusions are in Sec. \ref{conclusion}.

\section{Three-dimensional warped AdS black holes}
\label{background}
The models of gravity in three spacetime dimensions and their modifications have attracted a remarkable interest in the last decade. One of them is topologically massive gravity (TMG), which modifies the theory of general relativity (GR) by adding a Chern-Simons gravitational term \cite{Deser:1981wh} to the Hilbert-Einstein action. The action is described by 
\begin{equation}
I=\frac{1}{16 \pi G}\int_M d^3x\sqrt{-g}\left(R+\frac{2}{l^2}\right)+\frac{l}{96\pi G \nu}\int_M d^3x\sqrt{-g}\epsilon^{\lambda\mu\nu}\Gamma^r_{\lambda\sigma}\left(\partial_{\mu}\Gamma^\sigma_{r\nu}+\frac{2}{3}\Gamma^\sigma_{\mu\tau}\Gamma^\tau_{\nu r}\right)~,
\end{equation}
where $\epsilon^{\tau\sigma\mu}=1/\sqrt{-g}$ is the Levi-Civita tensor and $\nu$ is a dimensionless coupling constant, which is related to the graviton mass $\mu$ by $\nu=\mu l/3$. This model, in constrast to GR in three spacetime dimensions, has a propagating degree of freedom which corresponds to a massive graviton. Also, the possibility of constructing a chiral theory of gravity at a special point of the space of parameters \cite{Deser:1982vy} are some of its important characteristics. Accordingly, it was conjectured that a consistent quantum theory of the so-called chiral gravity can be defined at $\mu l=1$ or $\nu =1/3$ \cite{Li:2008dq}. However, for non-chiral values of $\mu l$, it was shown that exist other two warped AdS$_{3}$ vacuum solutions for every value of $\mu l \neq 3 $ \cite{Anninos:2008fx}.  The warped AdS$_{3}$ geometry corresponds to a one parameter stretched deformation of AdS$_3$. Further aspects of TMG can be found in \cite{Garbarz:2008qn, Nakasone:2009bn, Bergshoeff:2009aq, Oda:2009ys, Ohta:2011rv, Muneyuki:2012ur, Vasquez:2009mk} and references therein.\\

The metric describing the spacelike stretched black holes in ADM form is given by 
\begin{equation}\label{tl1}
ds^2=-N^2(r)dt^2+\ell ^2 R^2(r) \left(  d\phi + N_{\phi} (r) dt \right) ^2+\frac{\ell ^4 dr^2}{4 R^2 (r) N^2 (r)}~,
\end{equation}
where the metric functions are
\begin{equation}
R^2(r)= \frac{r}{4}\left(  3 (\nu^2-1)r+(\nu^2+3)(r_++r_-)-4 \nu \sqrt{r_+ r_-  (\nu^2+3)}\right)~,
\end{equation}
\begin{equation}
N^2 (r)= \frac{\ell ^2 (\nu^2+3)(r-r_+)(r-r_-)}{4 R^2}~,
\end{equation}
\begin{equation}
N_{\phi} (r)=\frac{2 \nu r-\sqrt{r_+ r_- (\nu^2+3)}}{2 R^2}~,
\end{equation}
being $\nu^{2} \geq 1$ the parameter that controls the stretching deformation, $r_{+}$ and $r_{-}$ are the outer and inner horizon, respectively. For $\nu= 1$ the metric reduces to the metric of BTZ black holes in a rotating frame.
An important feature of these black holes is that the Killing vector $\partial_{t}$ is spacelike everywhere in the spacetime and consequently its ergoregion extends to infinity; therefore, observers can not follow the orbits of $\partial_{t}$ in the exterior region. Also, the energy of a particle can have negative energy in the exterior region. It is worth to mention that the warped AdS$_{3}$ space also arises in other contexts, for instance see  \cite{Gurses:1994bjn, Rooman:1998xf, Duff:1998cr, Israel:2003ry, Andrade:2005ur, Bengtsson:2005zj, Banados:2005da, Son:2008ye, Balasubramanian:2008dm, Giribet:2015afn}. 

\section{Motion of particles in the three-dimensional warped AdS black hole background}
\label{motion}
The equations of the geodesics can be derived from the Lagrangian of a test particle, which is given by \cite{chandra}
\begin{equation}
\mathcal{L}=-\frac{1}{2}\left(g_{\mu\nu}\frac{dx^\mu}{d\tau}\frac{dx^\nu}{d\tau} \right)~.
\end{equation}
So, for the three-dimensional warped AdS black hole (\ref{tl1}), the Lagrangian reads 
\begin{equation}\label{tl4}
  2\mathcal{L}=-\left(-N^2(r)+\ell^2 R^2(r) N_{\phi}^2 (r)\right)\dot{t}^2+2 \ell^2 R^2(r) N_{\phi}(r)\dot{t}\dot{\phi}+
 \frac{\ell^4}{4 R^2 (r) N^2 (r)}\dot{r}^2+ \ell^2 R^2(r)\dot{\phi}^2~,
\end{equation}
where $\dot{q}=dq/d\tau$, and $\tau$ is an affine parameter along the geodesic that we choose as the proper time. Since the Lagrangian (\ref{tl4}) is
independent of the cyclic coordinates ($t,\phi$), then their
conjugate momenta ($\Pi_t, \Pi_{\phi}$) are conserved. The equations of motion are obtained from
$ \dot{\Pi}_{q} - \frac{\partial \mathcal{L}}{\partial q} = 0$, where $\Pi_{q} = \partial \mathcal{L}/\partial \dot{q}$
are the conjugate momenta to the coordinate $q$, and are given by
\begin{equation}
\Pi_{t} = \left(-N^2(r)+\ell^2 R^2(r) N_{\phi}^2 (r)\right)\dot{t}+ \ell^2 R^2(r) N_{\phi}(r)\dot{\phi}\equiv -E~,
\end{equation}
\begin{equation}
\Pi_{r}= \frac{\ell^4}{4 R^2 (r) N^2 (r)}\dot{r}~,
\end{equation}
\begin{equation}
 \Pi_{\phi}
=  \ell^2 R^2(r) N_{\phi}(r)\dot{t} +\ell^2 R^2(r)\dot{\phi}\equiv L~,
\label{w.11c}
\end{equation}
where $E$ and $L$ are dimensionless integration constants associated to each of them.
The Hamiltonian is given by
\begin{equation}
\mathcal{H}=\Pi_{t} \dot{t} + \Pi_{\phi}\dot{\phi}+\Pi_{r}\dot{r}
-\mathcal{L}~,
\end{equation}
\begin{equation}
2\mathcal{H}=-E\, \dot{t} + L\,\dot{\phi}+\frac{\ell^4}{4 R^2 (r) N^2 (r)}\dot{r}^2\equiv -m^2~.
\label{w.11z}
\end{equation}
By normalization, we shall consider that $m^2 = 1$ for massive particles ($m^2 = 0$ for photons).
We solve the above equation for $\dot{r}^2$ in order to obtain the radial equation, $\dot{t}$, and $\dot{\phi}$, which allows us to
characterize the possible movements of the test particles without an explicit solution of the equations of motion, which yields
\begin{eqnarray}
&&\dot{t}= \frac{E+LN_{\phi}}{N^2}~,\\
\label{w.12}
&&\dot{\phi}= -\frac{ E N_{\phi}}{N^2}+L\left( \frac{1}{\ell^2 R^2}-\frac{N_{\phi}^2}{N^2} \right)~,\\
\label{w.13}
&&\dot{r}^{2}=  \frac{4 N^2 R^2 }{\ell^4} \left(\frac{1}{N^2} (E+L N_{\phi})^2-\frac{L^2}{\ell^2 R^2}-1    \right)~,
\label{rdot}
\end{eqnarray}
where the above equations represent all nonzero 4-velocity components $u=(\dot{t},\dot{r},\dot{\phi})$ for the geodesic motion that will be used in the next section to obtain the CM energy of two colliding particles falling freely from rest with the same rest mass $m_{0}$ in the warped AdS$_{3}$ black hole background.
We will assume $\dot{t}>0$ for all $r>r_+$ so that the motion is forward in time outside the horizon. So, the following condition must be fulfilled 
\begin{equation}
\label{condition}
E+LN_{\phi}>0, \,\, \text{for all}\,\, r>r_+~.
\end{equation}
Now, in order to see if a particle can reach the event horizon, we will analyze the effective potential of the three-dimensional warped AdS black hole by using the equation of motion of the particle in the radial direction given by
\begin{equation}\label{pot}
\dot{r}^{2}+V=0 \,,
\end{equation}
where $V$ is the effective potential of the particle in the radial direction, hence by comparing Eq. (\ref{rdot}) with Eq. (\ref{pot}), we obtain:
\begin{equation}
V(r)=-\frac{4}{l^{4}}R^{2}(r)\left( \left( E+LN_{\phi}(r)\right) ^{2}-N^{2}(r)\left(1+\frac{L^{2}}{l^{2}R^{2}(r)}\right)\right)~.
\end{equation}
The motion of the particle is allowed in regions where $V(r)\leq 0$ and it is prohibited in regions where $V(r)>0$. It is clear that the particle can exist on the event horizon $r=r_{+}$ because $N^{2}(r_{+})=0$, and then the effective potential is negative.
On the other hand, when $r \rightarrow \infty$ it is easy to show that the effective potential is giving by
\begin{equation}
V(r \rightarrow \infty) \approx \frac{((\nu^{2}+3) \ell ^2-3 E^{2}(\nu^{2}-1))r^{2}}{l^{4}}~.
\end{equation}
This expression shows that the existence of massive particles at infinity depends on the warped parameter $\nu$ and on its energy $E$. Therefore, massive particles can exist at infinity when the following condition is fulfilled, with $\nu \neq 1$ 
\begin{equation}\label{aa}
E^2>\frac{(\nu^2+3) \ell^2}{3(\nu^2-1)}\, .
\end{equation}

\begin{figure}[!h]
\begin{center}
\includegraphics[width=80mm]{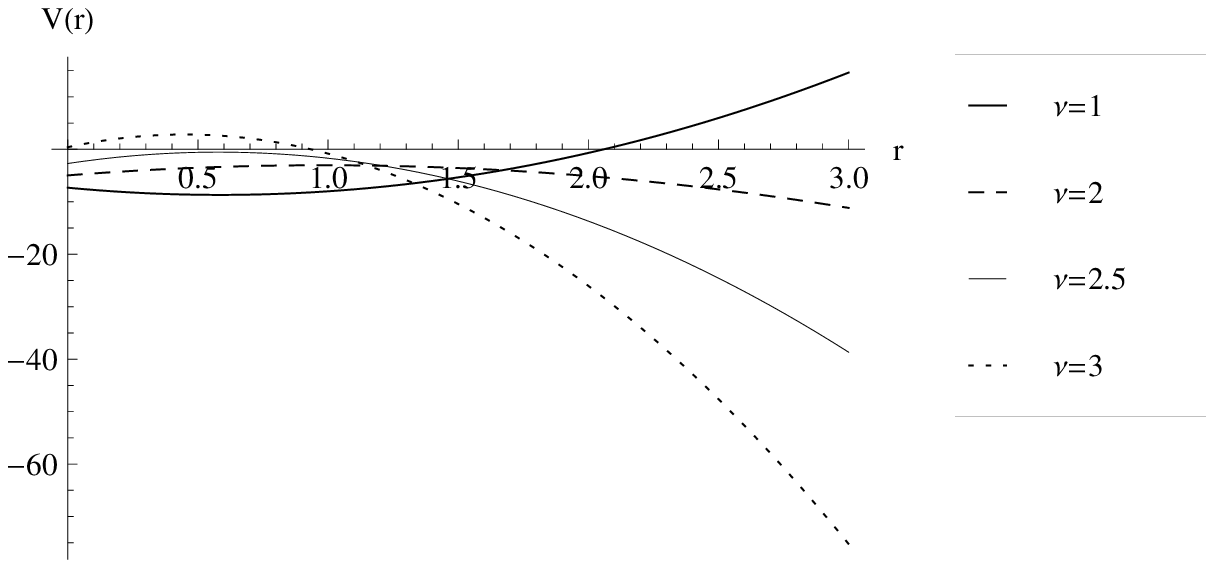}
\includegraphics[width=80mm]{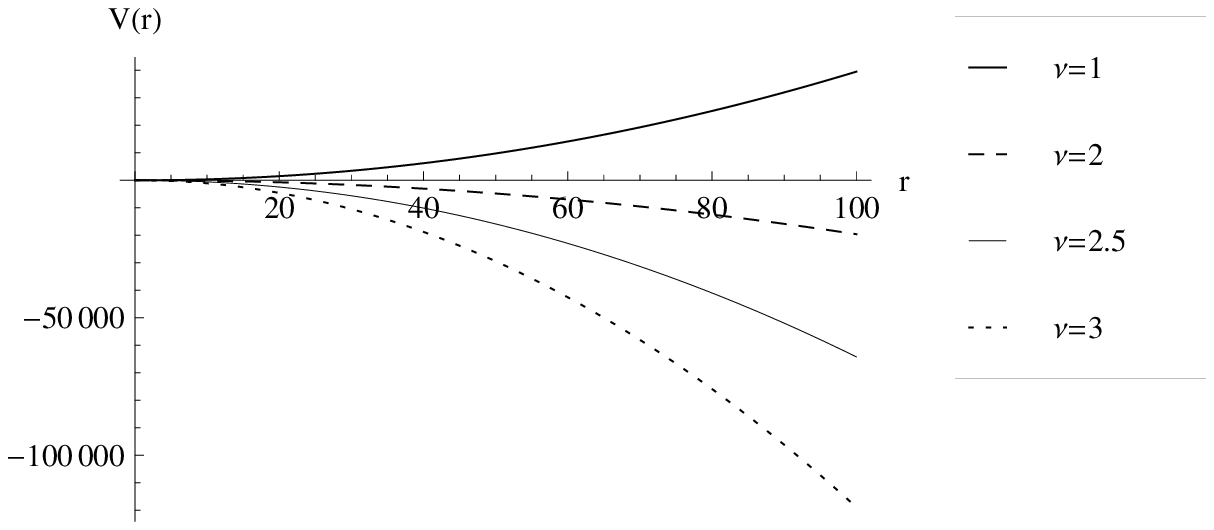}
\end{center}
\caption{The behavior of the effective potential $V(r)$ as a function of $r$ for different values of $\nu$ with $E=l=1$,$L=-1$, $r_-=1$, $r_+=2$ and $\nu=1,2,2.5,3$. Left figure shows the behavior of the effective potential between $r=0$ and $r=3$, while that the right figure shows the behavior of the effective potential between $r=0$ and $r=100$.}
\label{f1}
\end{figure}

\begin{figure}[!h]
\begin{center}
\includegraphics[width=80mm]{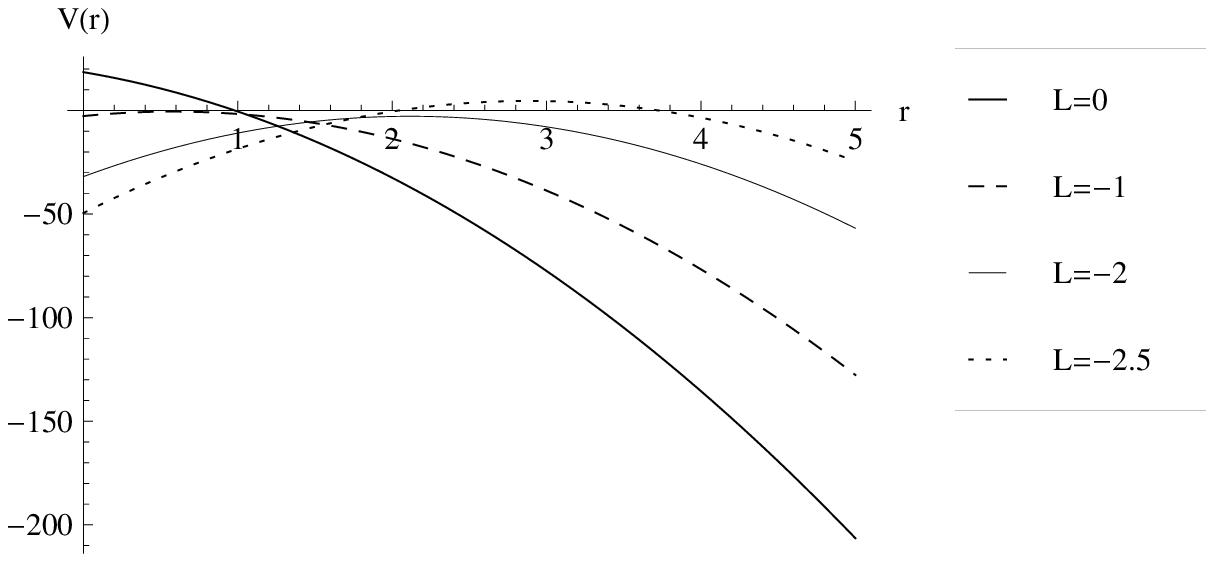}
\includegraphics[width=80mm]{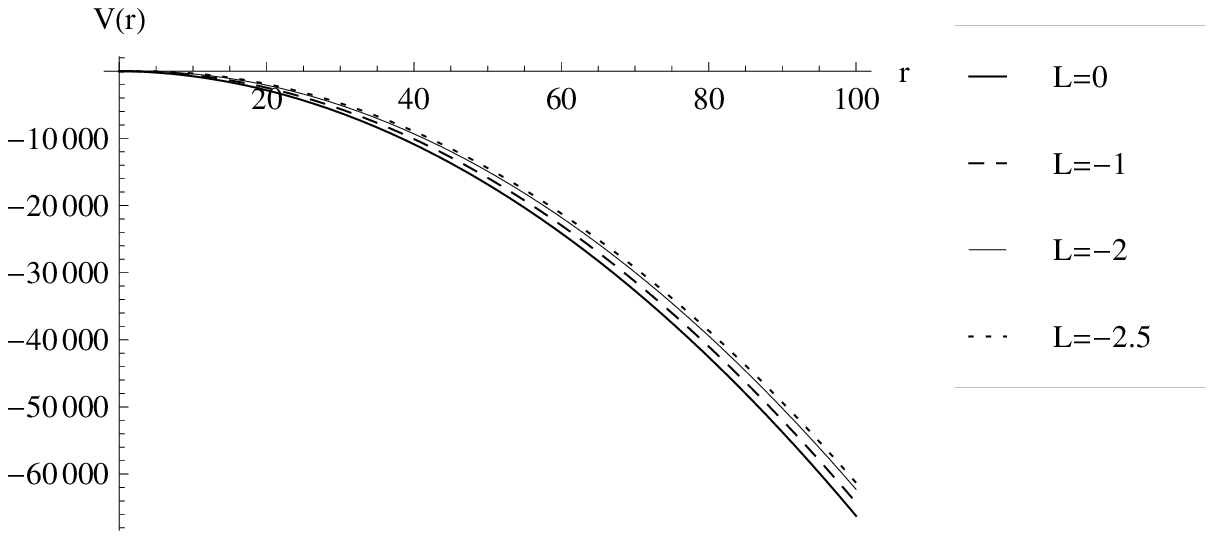}
\end{center}
\caption{The behavior of the effective potential $V(r)$ as a function of $r$ for different values of $L$ with $E=l=1$, $r_-=1$, $r_+=2$, $\nu=2.5$ and $L=0,-1,-2,-2.5$. Left figure shows the behavior of the effective potential between $r=0$ and $r=5$, while that the right figure shows the behavior of the effective potential between $r=0$ and $r=100$.}
\label{f1.1}
\end{figure}

For $\nu = 1$, which corresponds to the BTZ black hole in a rotating frame, the effective potential is positive and does not depend on the energy; therefore, a massive particle can not exist at infinity. We plot the effective potential as a function of $r$ in Fig. \ref{f1}, for some values of the parameters that satisfy Eq. (\ref{condition}), and it can be seen that massive particles with positive energy and $\nu> 1$ can exist outside the event horizon $r\geq r_{+}$, and reach it, by depending on the angular momentum $L$ of the particle, see Fig. \ref{f1.1}. 
As we pointed out, for the special case $\nu=1$, which correspond to BTZ black hole in the rotating frame, the massive particles can not exist at infinity, which is similar to the behavior found in \cite{Tsukamoto:2017rrl} and \cite{Yang:2012we} for the BTZ black hole.

\newpage

\section{The CM energy of two colliding particles}
\label{CMS}
\begin{figure}[!h]
\begin{center}
\includegraphics[width=80mm]{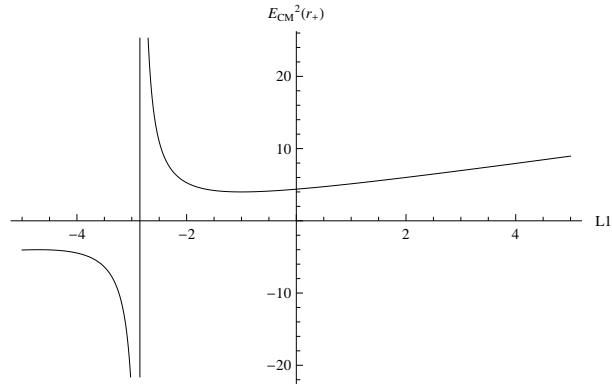}
\end{center}
\caption{The behavior of the CM energy $E_{CM}^2(r_+)$ at the horizon as a function of  $L_1$ with $\nu=2.5$, $E_1=E_2=1$, $L_2=-1$ $r_-=1$, $r_+=2$ and $l=1$.}
\label{f3}
\end{figure}

\begin{figure}[!h]
\begin{center}
\includegraphics[width=80mm]{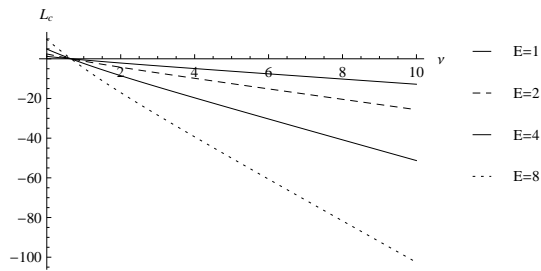}
\end{center}
\caption{The behavior of $L_{c}$ as a function of  $\nu$  for different values of the energy $E=1,2,4,8$ with $r_-=1$ and $r_+=2$.}
\label{f4}
\end{figure}

In this section we calculate the CM energy of two colliding particles in the warped AdS$_{3}$ black hole. To achieve this, we have to derive the 2+1 dimensional 4-velocity components to obtain the CM energy of the colliding particles. We consider that the particles have the same rest mass $m_0$, energies $E_1$ and $E_2$ and angular momentum $L_1$ and $L_2$, respectively. From the relation $E_{CM}=\sqrt{2} m_0 \sqrt{1-g_{\mu \nu}u_1^{\mu}u_2^{\nu}}$, where $u_1$ and $u_2$ denotes the 4-velocities of the particles, we obtain
\begin{equation}\label{CM}
\frac{E_{CM}^2}{2 m_0^{2}}= \frac{R^2 N^2 (R^2 -L_1 L_2 / \ell^2 )+(K_1 K_2-H_1 H_2)}{(R^2)^2 N^2}~,
\end{equation}
where
\begin{eqnarray}
\notag K_i  &=& E_i R^2+ L_i R^2 N_{\phi}~, \\
H_i &=& \sqrt{(E_i R^2+ L_i R^2 N_{\phi})^2-R^2 N^2 (R^2+L_i^2 /\ell^2)}~,
\end{eqnarray}
being $i=1,2$. Also, when the particles arrive to the horizon $r=r_{+}$, $N^2(r_+) \rightarrow 0$, $H_1 \rightarrow \sqrt{K_1^2}$ and $H_2 \rightarrow \sqrt{K_2^2}$, the CM energy (\ref{CM}) at the horizon yields:
\begin{equation}
\frac{E_{CM}^2}{2 m_0^{2}}(r \rightarrow r_{+})=\frac{1}{\ell^2 r(R^2)^2 N^2} (K_1 K_2 -\sqrt{K_1^2} \sqrt{K_2^2})~.
\end{equation}
Note that for $K_1K_2<0$ the $E_{CM}$ on the horizon ($N^2(r_+)=0$) is infinite negative, which is not a physical solution. However, when $K_1K_2 >0$, the numerator of  this expression will be zero and the value of $E_{CM}$ will be undetermined. Now, in order to find the limiting value of the CM energy at the horizon we use the L'Hopital rule, obtaining
\begin{equation}
\frac{E_{CM}^2}{2 m_0}=\frac{ R^2(r_+) (K_1(r_+)+K_2(r_+))^2 +\frac{1}{\ell^2}(K_1(r_+) L_2-K_2(r_+)L_1)^2}{2 R^2(r_+) K_1(r_+) K_2(r_+)}~.
\end{equation}
Note that the numerator of the above expression is finite at the horizon and when $K_i(r_+)=0$ the CM energy of two colliding particles on the horizon could be arbitrary high $E_{CM} \big | _{K_i=0} \rightarrow \infty$. So, from $K_i(r_+)=0$ we obtain that the critical angular momentum is given by:
\begin{equation}\label{critico}
L_{ci}= \frac{E_i r_+ \left( 3r_-+\nu^2 r_-+4\nu^2 r_+-4 \nu \sqrt{r_+ r_- (3+\nu^2)} \right)}{2 \left( -2 \nu r_++\sqrt{r_+ r_- (3+\nu^2)} \right)}~,  \,\,\,\,\, i=1,2.
\end{equation}
On the other hand, when $K_1(r_+)$ and $K_2(r_+)$ are both zero, then $E_{CM}$ is finite at the horizon. In this case $H_1(r_+)=H_2(r_+)=0$ and
\begin{equation}
\frac{E_{CM}^2}{2 m_0}=1-\frac{L_1 L_2}{ \ell^2 R^2(r_+)}~.
\end{equation}
Therefore, in order to obtain an infinite CM energy only one of the colliding particles should have the critical angular momentum, being the BSW process possible.
In Fig. \ref{f3}, we show the behavior of $E_{CM}$ versus $L_{1}$ with the other parameters fixed. We observe that exist a critical value of angular momentum for the particle 1 for which the CM energy blows up. Note that in order to get positive $E_{CM}$ the asymptotic value of critical angular momentum has to be reach from the right. Additionally, in Fig. \ref{f4}, we have plotted $L_{c}$ in terms of warped parameters $\nu$ for different values of energy $E$. Notice that when the energy of the particle 1 and the warped parameter increases, the critical angular momentum $L_{c}$ decreases.
By a similar analysis it is possible to evaluate the $E_{CM}$  on the inner horizon, finding that this is also infinity as long as  one of the two particles have the following  critical angular momentum
\begin{equation}\label{critic}
L_{ci}=\frac{E_i r_- \left(3r_++\nu^2 r_+ +4\nu^2 r_- -4\nu \sqrt{r_-r_+(3+\nu^2)}\right)}{2 \left( -2 \nu r_-+\sqrt{r_-r_+(3+\nu^2)}\right)}~, \,\,\,\,\, i=1,2.
\end{equation}

\newpage

\section{Radial motion of the particle with critical angular momentum}
\label{two}
\begin{figure}[!h]
\begin{center}
\includegraphics[width=70mm]{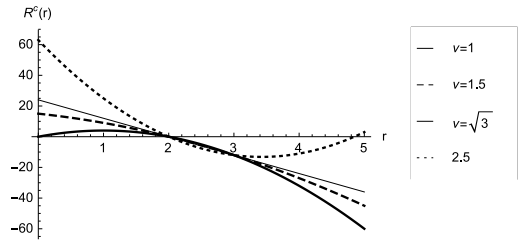}
\end{center}
\caption{The behavior of $R^c(r)$ as a function of $r$ with $r_- = 1$, $r_+=2$, $\nu=1\, (L_c=-0.586)$, $\nu=1.5\, (L_c=-1.380)$, $\nu=\sqrt{3}\, (L_c=-1.732)$, $\nu=2.5\, (L_c=-2.894)$, $E=1$ and $l=1$.}
\label{f5}
\end{figure}

In this section, we will study the radial motion of the particle with critical angular momentum and energy $E$. As we have mentioned, the particle reaches the event horizon of the black holes if the square of the radial component of the 4-velocity $\dot{r}^{2}$ in Eq. (\ref{rdot}) is positive or $V$ is negative in the neighborhood of the black hole horizon. We will denote the explicit form of $\dot{r}^{2}$ with critical angular momentum as $R^{c}(r)$, which is given by
\begin{equation}\label{POTC}
R^{c}=\frac{(r-r_{+})\left(-m^{2}l^{2}(r-r_{-})(3+\nu^{2})+E^{2}(3r(-1+\nu^{2})+r_{-}(3+\nu^{2})-4r_{+}\nu^{2})\right)}{l^{4}}~,
\end{equation}
and it vanishes on the event horizon. Also, for some values of the parameters, $R^c$ can be positive, which implies that particles with critical angular momentum can exist outside the event horizon, but as we shall see, they cannot reach the event horizon, unless $r_+=r_-$.
Particles with critical angular momentum can reach the event horizon if
\begin{equation}
\frac{dR^{c}}{dr}\Big|_{r=r_{+}}>0\, ,
\end{equation}
which yields
\begin{equation}
\frac{dR^{c}}{dr}\Big|_{r=r_{+}}=-\frac{(E^{2}+l^{2}m^{2})(r_{+}-r_{-})(\nu^{2}+3)}{l^{4}}<0~.
\end{equation}
Therefore, if particles with critical angular momentum exist outside the black hole cannot reach the event horizon. In Fig. \ref{f5} we plot the behavior of $R^c$ as a function of $r$ for the three-dimensional warped black hole. Notice that if $\frac{d^{2}R^{c}}{dr^{2}}=\frac{-2\left( m^{2}l^{2}(3+\nu^{2}\right)-3E^{2}(\nu^{2}-1))}{l^{4}}>0$, 
$R^c(r)$ has a zero also at $r_0=\frac{(\nu^2+3) \ell^2 r_-+ (3 r_-+(r_- -4 r_+)\nu^2) E^2 }{(\nu^2+3)\ell ^2 -3 E^2 ( \nu^2-1)}$ which is greater than $r_+$, and particles with critical angular momentum can exist outside the event horizon at $r>r_0 >r_+$, see Fig. \ref{f5}.

 On the other hand, we notice that the particle with the critical angular momentum can exist inside the event horizon $r_{+}$. This can be shown by replacing the critical angular momentum Eq. (\ref{critic}) into the square of the radial component of the 4-velocity Eq. (\ref{rdot}), obtaining the analog of Eq. (\ref{POTC}), whose derivative evaluated on the internal horizon $r_-$ is positive $\frac{dR^{c}}{dr}\vert_{r=r_-}>0$; therefore, the particle with critical angular momentum can to reach the inner  horizon and the center of mass energy can be arbitrarily high, being possible the BSW process on the inner horizon.

\begin{figure}[!h]
\begin{center}
\includegraphics[width=90mm]{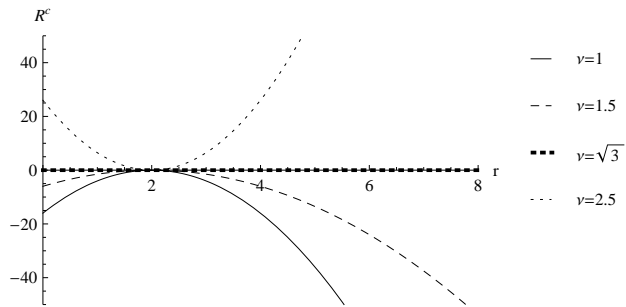}
\end{center}
\caption{The behavior of $R^c(r)$ as a function of $r$ for the extremal warped AdS$_3$ black hole for different values of  the warped parameter $\nu=1, \nu=1.5\, (L_c=-0.709),\, \nu=\sqrt{3}\, (L_c=-1.015), \nu=2.5\, (L_c=-1.959)$, $E=1$ and $l=1$.}
\label{f6}
\end{figure}

On the other hand, in the extremal case $r_{+}=r_{-}$ we obtain
\begin{equation}\label{extremal}
R^{c}=-\frac{(r-r_{+})^{2}\left( m^{2}l^{2}(3+\nu^{2})-3E^{2}(\nu^{2}-1)\right) }{l^{4}}~,
\end{equation}
and
\begin{equation}\label{dextremal}
\frac{dR^{c}}{dr}=-\frac{2(r-r_{+})\left( m^{2}l^{2}(3+\nu^{2})-3E^{2}(\nu^{2}-1)\right) }{l^{4}}~.
\end{equation}
Then, clearly Eqs.(\ref{extremal}) and (\ref{dextremal}) are zero on the event horizon, then it is necessary to calculate $\frac{d^{2}R^{c}}{dr^{2}}\vert_{r=r_{+}}$:
\begin{equation}
\frac{d^{2}R^{c}}{dr^{2}}\Big|_{r=r_{+}}=\frac{-2\left( m^{2}l^{2}(3+\nu^{2}\right)-3E^{2}(\nu^{2}-1))}{l^{4}}~.
\end{equation}
If $\frac{d^{2}R^{c}}{dr^{2}}\Big|_{r=r_{+}}>0$, the particle with critical angular momentum will reach the degenerate horizon, in our case it is fulfilled if:
\begin{equation}\label{con1}
E^2> \frac{(\nu^2+3) \ell^2}{3 (\nu ^2 - 1 )}\, ,
\end{equation}
which is the same expression given in Eq. (\ref{aa}). 
In Fig. \ref{f6} we plot the behavior of the $R^c$ as a function of $r$ for the  three-dimensional extremal warped AdS black hole.
Notice that 
the particle with critical angular momentum can reach the degenerate horizon if the condition (\ref{con1}) 
is satisfied, being the BSW process possible. This result is similar to that found in \cite{Li:2010ej} for the Kerr-AdS black holes. Notice that for $\nu =1$ the particle with critical angular momentum cannot exist outside the event horizon; however can exist on the degenerate horizon, which is similar to the behavior for the extremal BTZ black hole \cite{Yang:2012we}.

\newpage

\section{Final remarks}
\label{conclusion}

In this paper we considered two colliding particles in the vicinity of the horizon of a three-dimensional warped AdS black hole, and we analyzed the energy in the CM frame of the colliding particles in order to investigate the possibility that 
the three-dimensional warped AdS black hole can act as a particle accelerator.
We found for the warped AdS$_3$ black hole that despite the particle with critical angular momentum can exist for certain values of the conserved energy outside the black hole it will never reach the event horizon; therefore, the black hole can not act as a particle accelerator with arbitrarily high CM energy on the event horizon. However, the particle with the critical angular momentum could also exist inside the outer horizon of the warped AdS$_3$ black hole being the BSW process possible on the inner horizon. Also, we shown that the critical angular momentum decreases when the parameter that controls the stretching deformation increases.
At $\nu=1$, the black hole solution is given by the 
BTZ black hole in the rotating frame, and we shown that the massive particles can not exist at infinity, which is similar to the behavior observed for  
the BTZ black hole.

On the other hand, for the extremal warped AdS$_{3}$ black holes, we found that the particle with critical angular momentum could exist outside the event horizon and reach a high CM energy on the event horizon as long as its conserved energy fulfill the condition $E^{2}>\frac{(\nu^{2}+3)l^{2}}{3(\nu^{2}-1)}$, being the BSW process possible. At $\nu =1$ the particle with critical angular momentum cannot exist outside the event horizon; however can exist on the degenerate horizon, which is similar to the behavior for the extremal BTZ black hole.

\acknowledgments

This work was partially funded by the Comisi\'{o}n
Nacional de Ciencias y Tecnolog\'{i}a through FONDECYT Grant 11140674 (PAG) and by the Direcci\'{o}n de Investigaci\'{o}n y Desarrollo de la Universidad de La Serena (Y.V.). P. A. G. acknowledges the hospitality of the Universidad de La Serena, National Technical University of Athens and Pontificia Universidad Cat\'{o}lica de Valpara\'{i}so where part of this work was undertaken.

\appendix

\end{document}